\documentclass[twocolumn,showpacs,preprintnumbers,amsmath,amssymb,prb,floatfix]{revtex4}

\usepackage{graphicx}
\usepackage{dcolumn}
\usepackage{bm}
\usepackage{epstopdf} 
 
\newcommand{\be}{\begin{equation}} 
\newcommand{\ee}{\end{equation}}
\newcommand{\bea}{\begin{eqnarray}}
\newcommand{\eea}{\end{eqnarray}}
\begin{document}
\bibliographystyle{prsty}
\title{Photoinduced magnetism in the ferromagnetic
semiconductors }
\author{Subodha Mishra and Sashi Satpathy}
\address {Department of Physics \& Astronomy,
 University of Missouri, Columbia, MO 65211, USA}
\date{\today}
\begin{abstract}
We study the enhancement of the magnetic transition temperature $T_c$
due to incident light in ferromagnetic semiconductors 
such as EuS.
The photoexcited carriers mediate an extra ferromagnetic interaction
due to the coupling with the localized magnetic moments. 
The Hamiltonian consists of a Heisenberg model for the 
localized moments and an interaction term between
the photoexcited carriers and the localized moments. The
model predicts a small enhancement of the transition temperature in semi-quantitative agreement with the experiments.
\end{abstract}
\maketitle

There is a class of materials which are ferromagnetic 
semiconductors\cite{nolt,paul,lems,sanford,rudov} 
such as the  Europium 
chalcogenides EuX (X=S, O, Se, Te) and 
chalcogenide spinels MCr$_2$Y$_4$  
 (M=Hg, Cd; Y=S,Se). The ferromagnetism in these materials are generally well 
described by the Heisenberg model. In this paper, we are concerned with 
the effect of incident light on the magnetism, in particular, on the ferromagnetic transition temperature $T_c$. 
 The basic mechanism of the photoenhanced magnetism is that light excites 
electron-hole pairs, 
which mediate an extra magnetic interaction between the local magnetic moments, on top of any pre-existing magnetic interaction, thereby enhancing the $T_c$. The related phenomena, where ferromagnetism occurs only if light is present but otherwise the system is paramagnetic, has been studied in connection with the dilute magnetic semiconductors.\cite{krenn,koshi,oiwa,mitsu,furdyna,mis,subodha,ferna}

Here we study the photoinduced magnetism in the ferromagnetic semiconductors by solving a model Hamiltonian
within a mean-field approach by constructing a BCS-type wave function for the electron-hole pairs,\cite{bcs} similar to our analysis  developed earlier for the dilute magnetic semiconductors.\cite{mis}  Our work  provides a semi-quantitative explanation for the enhancement of $T_c$, which has been observed to be as large as $0.16 K$ for EuS\cite{naga,afa1,afa2} 
under intense laser radiation.  
\begin{figure}[ht]
\includegraphics[angle=0,width=0.75\linewidth] {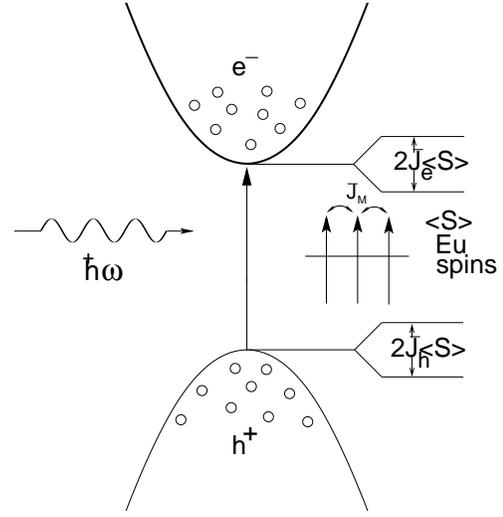}
\caption{
A schematic picture of the photoinduced ferromagnetism in EuS. 
Incident light creates electron-hole pairs that mediate an extra ferromagnetic exchange interaction between the Eu moments. The net result is an enhancement of the critical temperature $T_c$ by a fraction of a degree K for typical parameters.
}
\label{bands}
\end{figure} 

The ferromagnetic semiconductor  is modeled by 
isotropic conduction and valence bands interacting with fixed
Eu magnetic moments originating from the f electrons. The bands are spin-split 
through an exchange interaction with the magnetic ions
as indicated in Fig. \ref{bands}. The Hamiltonian
consists of a Heisenberg interaction term between the fixed Eu magnetic moments  and  a term describing the itinerant band electrons
and their interaction with light.
We thus write
\be
{\cal H}= -\sum_{<ij>} J_M S_i.S_j + {\cal H}_{eh}^L,
\ee
where the first term is the Heisenberg term  and $<ij>$ denotes summation over
distinct pairs of nearest neighbours. Each Eu atom has twelve nearest neighbor in the fcc lattice of EuS.
The second term ${\cal H}_{eh}^L$ describes electrons and holes 
making up the conduction and valence bands and their 
interaction with light
\be
{\cal H}_{eh}^L={\cal H}_{kin}+ {\cal H}_c+ {\cal H}_L(t)\label{hhh}.
\ee
The kinetic energy term is given by
\be
{\cal H}_{kin}=\sum_{k\sigma}(E^e_{k\sigma}
c_{k\sigma}^\dagger c_{  k\sigma} 
+ E^h_{k\sigma}d_{  k\sigma}^\dagger d_{  k\sigma}),
 \label{hkin}
\ee
where $c_{k\sigma}^\dagger, c_{  k\sigma} $ and $d_{k\sigma}^\dagger, d_{  k\sigma} $ are the
field operators for the electrons and holes, respectively, with $k$ being the Bloch momentum and $\sigma$, the spin index. Throughout this paper, electrons refer to the photoexcited electrons in the conduction bands and holes refer to the lack of an electron in the valence bands. Since the photon has zero spin and negligible momentum, absorption of light involves a vertical transition of an electron in the valence bands into the conduction bands with the spin of the electron conserved thus resulting in the creation of an electron-hole pair described by $c_{  k\sigma}^\dagger d_{-  k -\sigma}^\dagger$. Without incident light, we have the ``vacuum state" with no electrons nor holes.

The single-particle energies in Eq. \ref{hkin}
include the Zeeman splitting due to the electron interaction with the localized Eu spins, so that
\be
E^e_{k\sigma}=E_g+\frac{\hbar^2 k^2}{2m_e} \pm J_e  <S> 
\label{ec}
\ee
and
\be
E^h_{k\sigma}=\frac{\hbar^2 k^2}{2m_h} \pm J_h  <S> ,
\label{ev}
\ee
where $\pm$ describes the Zeeman splitting of the up-spin and the down-spin 
states. 
Here
$J_e$ and $J_h$ are the electron and hole exchange interactions with the 
localized Eu moments, $<M>=g \mu_B <S>$ is the average moment of a Eu 
atom, $m_e$ 
and $m_h$ 
are the electron and hole masses, respectively, and $E_g$ is the band 
gap. 

The Coulomb interaction part consists of three terms
\be
 {\cal H}_c= {\cal H}_{ee} + {\cal H}_{hh} +  {\cal H}_{eh},
\ee
  where the electron-electron interaction term is
\be
{\cal H}_{ee}=\frac{1}{2}\sum_{  k  k'  
q  \sigma\sigma'}V_{  q} \
c_{  k+  q, \sigma}^\dagger c_{  k'-  q, \sigma'}^\dagger c_{  
k'\sigma'}c_{  k\sigma},
\label{hee}
\ee
with a similar term for the hole-hole interaction, while the electron-hole interaction
is given by the expression
\be
{\cal H}_{eh}=-\sum_{  k  k'  q \sigma \sigma'}V_{  q} \
c_{  k+  q,\sigma}^\dagger d_{  k'-  q,\sigma'}^\dagger d_{ 
k'\sigma'}c_{  k\sigma}.
\label{heh}
\ee

Finally, the last term in Eq. (\ref{hhh}) which describes the coupling to the 
external electromagnetic 
field of 
frequency $\omega$,
in the ``rotating wave approximation" is given as
\be
{\cal H}_L(t)=\lambda\sum_{  k}\rho_k[(
c_{  k\uparrow}^\dagger d_{-  k\downarrow}^\dagger 
+c_{  k\downarrow}^\dagger d_{-  k\uparrow}^\dagger) e^{-i\omega 
t}+h.c.].
\label{hlight}
\ee
As mentioned earlier, the light induces a vertical transition with no net spin change and  $\lambda  \rho_k$ is the 
coupling 
strength for the creation of such an electron-hole pair. 
The coupling strength
is given as $\lambda\rho_k=eE<\psi_{vk}|\hat e\cdot \vec 
p|\psi_{ck}>/m\omega$ where $\psi_{ck},\   \psi_{vk}$ are  
conduction and the valence state wave functions,  
E is the strength of the electric field and  $\hat e$ is the electric 
polarization vector and $\vec p$ is the momentum operator. The  matrix element may be computed once the wave 
functions are  known, but here, for simplicity, we neglect the momentum 
dependence of the matrix elements, so that $\rho_k=1$ and the parameter 
$\lambda$ indicates the  strength of the interaction.  
For strong laser radiation available in the laboratory, a broadly estimated value  used by earlier authors\cite{ferna,Comte1} is $\lambda \approx 
0.1\ eV$, which we use here in our numerical calculations.

The time-dependence of Eq. (\ref{hlight}) can be eliminated\cite{mis,Comte1} by  an appropriate unitary 
transformation $e^{-iS}{\cal H} e^{iS}$ where $S=-\frac{\omega t}{2}
\sum_{k\sigma}(c_{k\sigma}^\dagger c_{  k\sigma}+d_{  k\sigma}^\dagger d_{  
k\sigma})$
and the Hamiltonian becomes
\bea
\tilde {\cal H} &=& -\sum_{<i,j>} J_M S_i.S_j\nonumber \\ 
&&+ \sum_{  k\sigma}  
(E^e_{k\sigma}
-\hbar\omega/2)c_{  k\sigma}^\dagger
c_{  k\sigma}
+(E^h_{k\sigma}-\hbar\omega/2) d_{k\sigma}^\dagger d_{  k\sigma}\nonumber 
\\
&&+{\cal H}_c
+\lambda\sum_{  k}  \{ \rho_{  k} { {(c_{ 
k\uparrow}^\dagger d_{-  k\downarrow}^\dagger
+c_{  k\downarrow}^\dagger d_{- 
k\uparrow}^\dagger)+h.c.}  } \}.
 \label{htnond2}
\eea
Thus the time has been eliminated and the transformed Hamiltonian 
may be interpreted
as a ``quasi free energy" with the chemical potential for the 
creation of an electron-hole pair given by
$\mu = \hbar \omega$.
One can thus write the transformed Hamiltonian in the form
$
{\tilde {\cal H} } = {\cal H}_{QE} - \mu N,
$
where $ N = \sum_{k\sigma}c_{k\sigma}^\dagger 
c_{k\sigma}$  is the number of electron-hole pairs.

The Coulomb interaction term appearing 
in the Hamiltonian
(\ref{hee}) and (\ref{heh}) is given by $V_q = 4 \pi e^2 / (\epsilon 
q^2)$, 
but in our discussions below,
 we replace it, for the sake of simplicity, by a contact interaction term  
$V(\vec r) = V_0 \delta({\vec r})$
 without changing the qualitative physics. Its typical magnitude in semiconductors is $V_0 \approx 20$  eV.\AA$^3$.\cite{mis}
Typical value for the exchange parameter is:
 $J_M=2.19\times 10^{-5}\ eV$, which is calculated using the 
expression for the critical temperature for the Heisenberg Hamiltonian $k_BT_c=(\nu/3) J_MS(S+1)$. For EuS, $T_c=16K$, $S=7/2$, and the number of nearest neighbours $\nu=12$.
The magnitude of  $J=J_e-J_h=-0.036\ eV$ as obtained from estimating $J_e$ and $J_h$ from the splitting of 
the spin-up and 
the spin-down bands in the band structure of EuS.\cite{nolt} 

 We compute the free energy of the system by using a variational wave function similar to the  
 Bardeen-Cooper-Schrieffer (BCS) wave function with zero net 
momentum and
 spin. The variational wave function consists of a superposition of the electron-hole pairs \cite{Comte3,Comte4}
generated by the coupling to light
\be
|\Psi>=\prod_{k\sigma}
(u_{k\sigma}+v_{k\sigma}c^\dagger_{k\sigma}d^\dagger_{-k-\sigma})|0>,
\label{BCS-wave}
\ee
where $|0>$ is 
the
intrinsic vacuum state (filled valence and empty conduction bands),
$v_{k\sigma}$ is the probability amplitude for the creation of an electron-hole pair  with the electron momentum and spin  of
$k\sigma$ and a hole with the opposite momentum and spin, while $u_{k\sigma}$ is the
probability amplitude for the absence of such pair excitation.
The normalization condition is, as usual,
$u_{  k\sigma}^2+v_{  k\sigma}^2=1$,
and the coefficients $u_{k\sigma}$ and $v_{k\sigma}$ may be chosen as real.
In the present case the 
localized moments break the symmetry between the 
spin up and down electrons
 leading to a net spin polarization of the 
electron and holes. This in turn affects the net magnetic moment, which
then affects the energies of the particles through the Zeeman term, so that 
all quantities have to be determined in a self-consistent manner.

We now calculate the expectation value of the ``free energy" from Eq. (\ref{htnond2}), which  may be rewritten in terms of the 
gap parameter to yield\cite{mis}
\bea
F_0 &=& \sum_{k\sigma}[(\xi_{k\sigma}+\Omega_{\sigma}/2)v^2_{k\sigma} 
-(\Delta_{\sigma} u_{k\sigma}v_{k\sigma}/2
+ \lambda  u_{k\sigma}v_{k\sigma})]\nonumber\\
&+& \frac{\nu}{2}J_M<S>^2\times {\cal N},
\label{F0}
\eea 
where
\begin{equation}
\Delta_{\sigma}=2V_0\sum_ {k'}u_ {k'\sigma}v_{k'\sigma 
}+2\lambda  
\end{equation}
and
\begin{equation}
\Omega_{\sigma}=-2V_0\sum_{k'}v_{k'\sigma}^2.
\label{Delta-Omega}
\end{equation}
Here, $\nu$ is the number of nearest neighbour (12 for the fcc lattice)   and the factor of 1/2 in the last term of Eq. \ref{F0} appears to avoid double counting. 
 ${\cal N}$ is the  total  number of sites in the crystal 
and  $\xi_{  
k\sigma}$ is the energy of the electron-hole pair
excitation with a spin-up electron and a
spin-down hole measured with respect to the photon energy
\bea
\xi_{  k\sigma} &=& E^e_{k,\sigma} + E^h_{k,-\sigma} - \hbar 
\omega \nonumber   \\ 
&=&E_g + \frac{\hbar^2 k^2}{2 m^*} \pm J <S>
-\hbar\omega,
\label{xi-k-pm}
\eea
where
$J = J_e-J_h$ and the effective mass of the electron-hole pair $m^{*-1}=m_e^{-1}+m_h^{-1}$.
 Minimizing the free energy 
with respect to $v_{k\sigma}$, 
$\partial F_0/  \partial v_{k\sigma}=0 $, and using the normalization conditions 
$u_{  k\sigma}^2+v_{k\sigma}^2=1$, we get the occupation amplitudes $u_{k\sigma}$ and $v_{k\sigma}$, which read
\be
u_{ k\sigma}^2=\frac{1}{2} [1+\frac{\xi_{ k\sigma}+\Omega_{
\sigma}}{\sqrt{\Delta_{\sigma}^2+(\xi_{ k\sigma}+\Omega_{
\sigma})^2}}] \label{uk}
\ee
and
\be
v_{ k\sigma}^2=\frac{1}{2} [1-\frac{\xi_{k\sigma}+\Omega_{
\sigma}}{\sqrt{\Delta_{ \sigma}^2+(\xi_{ k\sigma}+\Omega_{
\sigma})^2}}] \label{vk}.
\ee


\begin{figure}[ht]
\includegraphics[angle=0,width=0.75\linewidth] {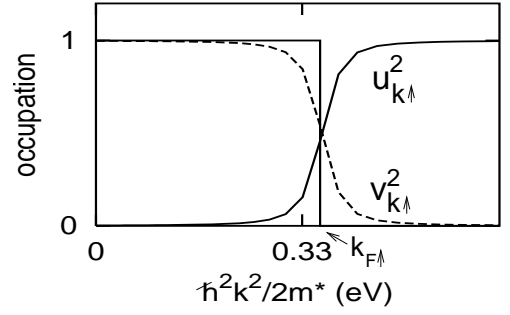}
\caption{Electron-hole pair amplitudes $u_{k\uparrow}$ and $v_{k\uparrow}$ as a function 
 of momentum k. Pairs of spin up electrons and spin down holes are filled up to
the Fermi momentum $k_{F\uparrow}$ without any interactions. The step function is 
in the limit of zero interactions ($V_0 = 0, \lambda \rightarrow 0$), while both the Coulomb 
interaction and the coupling to light  smear out the Fermi surface. 
Here parameters are: $\lambda = 0.1\ eV$,
 $m^* = 1.5 m_e$, $V_0 = 20$ eV.\AA$^3$ and $\hbar \omega - E_g = 0.33$ 
eV, $J_M=2.19\times 10^{-5}\ eV$ and $J=-0.036\ eV$. 
The pair amplitudes for the other spin, $u_{k\downarrow}$ and $v_{k\downarrow}$, (not shown in the figure)
are slightly shifted to the left, since $k_{F\downarrow}$ is slightly lower than 
$k_{F\uparrow}$.
 }
\label{ff155}
\end{figure}

We solve the equations (\ref{uk}) and (\ref{vk}) numerically which yield the results for
$u_{k\sigma}$ and $v_{k\sigma}$, from which the free energy $F_0$ is computed using Eq. (\ref{F0}). 
The typical pair-occupation amplitudes $u_k$ and $v_k$ have
 been shown in Fig. (\ref{ff155}). We note that in
the limit of small interaction
$(V=0$  and $\lambda=0)$, the free energy  
(Eq. (\ref{F0})) boils 
down to the energy of the noninteracting system of 
electron-hole pair states occupied up to the chemical 
potential (i.e., $v_{k\uparrow}^2$ 
and $\bar v_k{\downarrow}^2$ equal to one below $\hbar\omega$ and zero above).



Once the free energy is computed, the magnetization of the Eu moments is calculated within the
mean-field approximation. Each Eu moment experiences an effective
magnetic field H$_{eff}$ given by
\be
{\rm H}_{eff}=-\frac{\partial F_0(M)/{\cal N}}{\partial M}, 
\label{Heff}
\ee
so that the magnetization is given by the expression
\be
M(T)=g\mu_BSB_J(\beta g\mu_BSH_{eff}(M)),
\label{M}
\ee
where the B$_J$ is the Brillouin function
and 
 $\beta=(k_\beta T)^{-1}$ and $M(S) = g\mu_B S$.
The above two equations are solved self-consistently for a given temperature
using the free energy expression Eq. (\ref{F0}).

Note that in our formulation, we have neglected the thermal excitation of the electrons from the valence to the conduction bands, because the number of such carriers is much smaller than the number of the photogenerated carriers.  The latter number, obtained by integrating $v_{k\sigma}^2$ over all momentum, is typically $\sim 10^{-4}/ $ \AA$^2$, while the number of intrinsic thermal carriers is many orders of magnitude smaller,\cite{Kittel} when the semiconductor bandgap is as large as $\sim 1 eV$, as is the case for EuS.

\begin{figure}[ht]
\includegraphics[angle=0,width=0.90\linewidth] {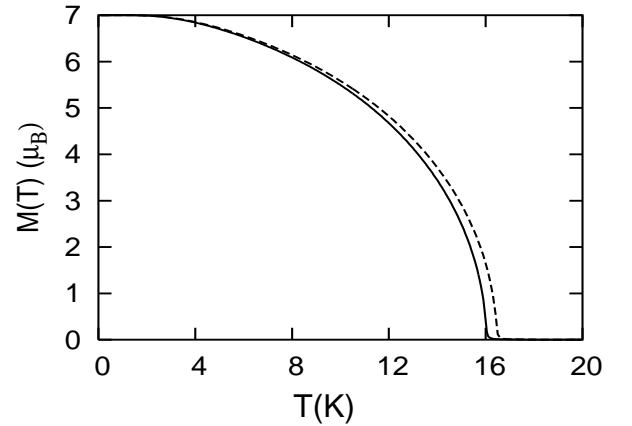}
\caption{
Dependence of the magnetization M 
 as a function of temperature T with (dashed line) and without (solid line) the incident light. Parameters are: $\lambda=0.1eV$, $m^*=1.5$, 
$V_0=20\ eV.\AA^3$, $\hbar\omega-E_g=1.5 eV$,   
$J_M=2.19\times 10^{-5}\ eV$, and $J=-0.036\ eV$. 
} \label{MvsT}
\end{figure}

\begin{figure}[ht]
\includegraphics[angle=0,width=0.90\linewidth] {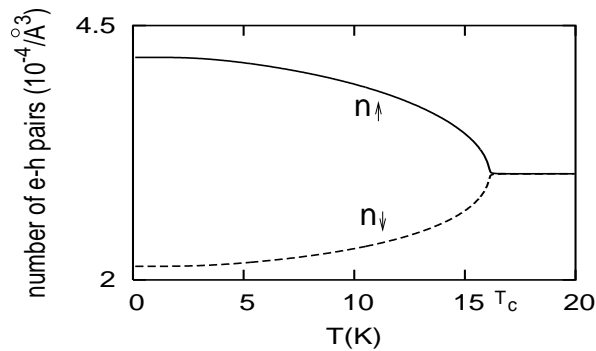}
\caption{
Pair occupancies as a function of temperature.
$n_{\uparrow}$ denotes the number of electron-hole pairs with
spin up electrons and spin down holes,
while $n_{\downarrow}$ denotes the number of pairs with
opposite spins. Below
the critical temperature,  $n_{\uparrow}$ is larger than $n_{\downarrow}$
because of the lower Zeeman energy of the former pairs. 
Above $T_c$, there is no net megnetization, so that the energy of the electron is independent of the spin (zero Zeeman splitting), so that the pair occupancies are independent of the spin of the electron or hole. Parameters are the
same as for Fig. \ref{MvsT}.}
\label{SvsT}
\end{figure}

\begin{figure}
\includegraphics[angle=0,width=0.90\linewidth] {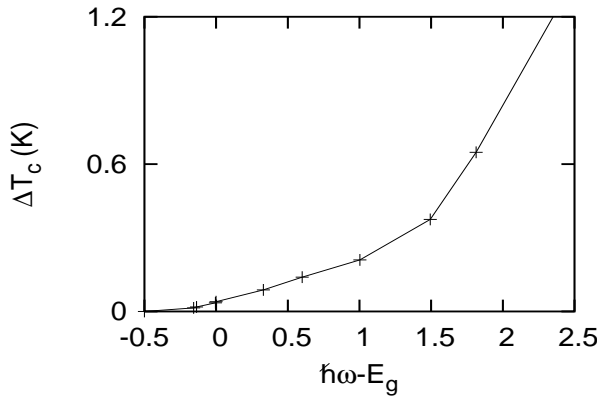}
\caption{
Photoinduced increase of the magnetic transition temerature as a function of the photon energy.  Parameters are the same as in Fig. 2. A larger photon energy leads to a larger number of carriers which add to the ferromagnet
}
\label{MvsT2}
\end{figure}

The calculated magnetization and the number of photoexcited electron-hole pair as a function of the temperature 
are shown in the
 Figs. \ref{MvsT} and {\ref{SvsT}} respectively. These are the two main results of this paper. As seen from the figures, there is a transition from
the paramagnetic to the ferromagnetic state as temperature is decreased. 
Fig. \ref{MvsT} shows that with the incident light and for typical parameters, the photoinduced increase in the transition temperature T$_c$ is of the order of 
a fraction of a degree, $\Delta T_c\sim 0.2K$, in line with 
the 
experimental value\cite{afa2} $\Delta T_c\sim 0.16K$ for the light frequency 
of
$\hbar\omega-E_g=0.33
eV$. 
 
Fig. {\ref{SvsT}} shows the number of pairs of spin-up electron and 
spin-down hole and pairs of spin-down electron and spin-up holes as 
function of temperature.
While the number of the electron-hole pairs is controlled by the chemical potential ($\mu =\hbar \omega$),
the distribution of the pairs between the two different spin types 
(spin-up electron and spin-down hole or vice versa) is determined by the magnitude of $\langle S \rangle$, 
which affects the
pair energy as indicated from Eqs. (\ref{xi-k-pm}) and (\ref{M}). The effective 
magnetic field experienced by the Eu spins is roughly proportional to 
the difference of the number of the two spin types, as may be seen by keeping the first line in the free energy Eq.
(\ref{F0}) and using Eqs. (\ref{xi-k-pm}) and (\ref{Heff}), so that $H_{eff} \propto 
(n_\uparrow - n_\downarrow)$. This in turn implies that this number difference 
and the magnetization M follow one another, both ultimately going to zero at the transition temperature T$_c$ as seen from Figs. \ref{MvsT} and {\ref{SvsT}}.
In the paramagnetic phase, the
photoexcited electron-hole pairs are present, but the two spin types are the same in number.

In Fig. {\ref{MvsT2} we plot the increase in $T_c$ vs. the incident light 
frequency, which as discussed earlier effectively serves as the chemical potential. 
We see that as the frequency increases, more and more 
photocarriers are created and hence $\Delta T_c$ increases. 
The figure shows that even when 
$\hbar\omega-E_g < 0$, there is an 
increase in $T_c$. This means that we have photoinduced 
magnetization even for the sub-bandgap light 
frequency, an issue that has been studied in detail by previous authors.\cite{ferna,Frohlich} Note that we have neglected the relaxation of the excited carriers, so that in our case carriers are filled up to the chemical potential $\mu=\hbar \omega$ at equilibrium. 
This relaxation effect 
will diminish the total number of photoexcited carriers, resulting in a somewhat lower value of $\Delta T_c$ than predicted from our analysis. Such relaxation effects must be taken into account if a more quantitative description is desired.


In summary, we studied the photoinduced enhancement of  magnetization 
in the 
ferromagnetic semiconductors 
by solving a model Hamiltonian which takes into account the Coulomb
interaction, light-matter interaction, the exchange
interaction between the carriers and the
fixed magnetic moments, as well as the direct exchange interaction among the localized
moments.
From the calculated free 
energy, we computed the 
magnetization of the system as a function of temperature. There is an  
enhancement of
$T_c$ by a fraction of a degree K due to carriers created by 
light, in semi-quantitative agreement with the observed experimental results for EuS.

This work was supported by the U. S. Department of Energy through Grant No.
DE-FG02-00ER45818.


\end{document}